\documentstyle[gcdv,psfig,natbib]{article}
\begin{document}

%%%%%%%%%%% Emulate PASA style %%%%%%%%%%%%%%%%%5
\small
\shorttitle{Physical Processes in Star-Gas Systems}
\shortauthor{R. Spurzem et al.}
%%%%%%%%%%% End Emulate PASA style %%%%%%%%%%%%%%%%%5

%
% Title
% Capitalise the title normally - do not use ALL CAPS.
%
\title
%%%%%%%%%%% Emulate PASA style %%%%%%%%%%%%%%%%%5
{\large \bf
%%%%%%%%%%% Emulate PASA style %%%%%%%%%%%%%%%%%5
%{
Physical Processes in Star-Gas Systems
}
%

% Authors
% Here comes the author(s) of the paper. Please add the appropriate author
% names for your paper and indicate within the $^...$ the number(s)
% which corresponds to the institute(s) of each author. In this example
% the second author has two institutional affiliations.
% Add or remove authors as required.
% **** IMPORTANT: Leave the closing curly bracket line as is. ******

%%%%%%%%%%% Emulate PASA style %%%%%%%%%%%%%%%%%5
\author{\small 
%%%%%%%%%%% Emulate PASA style %%%%%%%%%%%%%%%%%5
%\author{
 R. Spurzem $^1$, P. Berczik $^{1,2}$, G. Hensler $^{3}$, Ch. Theis  $^{3}$,
 P. Amaro-Seoane $^{1}$, M. Freitag $^{1}$, A. Just $^{1}$
} % IMPORTANT: leave this curly bracket as the first character of this line.

% Date - leave this blank.
\date{}
%%%%%%%%%%% Emulate PASA style %%%%%%%%%%%%%%%%%
\twocolumn[
%%%%%%%%%%% Emulate PASA style %%%%%%%%%%%%%%%%%5
\maketitle
\vspace{-20pt}
\small
% Institutions
% Here fill in your institute name(s) and address(es)
% The number in $^...$ indicates the author number.  For example
{\center
$^1$Astronomisches Rechen-Institut, M\"onchhofstr. 12-14, 69120 Heidelberg,
Germany\\
\{spurzem,pau,freitag,just\}@ari.uni-heidelberg.de\\
$^2$Main Astronomical Observatory of Ukrainian National Academy of Sciences,
Zabolotnoho Str. 27, 03680 Kiev, Ukraine \\
berczik@mao.kiev.ua \\
$^3$Institut f\"ur Theoretische Physik und Astrophysik, University of Kiel,
Olshausenstr. 40, 24098 Kiel, Germany \\
\{hensler,theis\}@astrophysik.uni-kiel.de\\[3mm]
}

% Abstract
% Simply place your abstract between the \begin{abstract} and
% \end{abstract} commands.
%
%\begin{abstract}
%%%%%%%%%%% Emulate PASA style %%%%%%%%%%%%%%%%%
\begin{center}
{\bfseries Abstract}
\end{center}
\begin{quotation}
\begin{small}
\vspace{-5pt}
%%%%%%%%%%% Emulate PASA style %%%%%%%%%%%%%%%%%
% Place the abstract here.
First we present a recently developed 3D
chemodynamical code for galaxy evolution from the K**2 collaboration.
It follows the
evolution of all components of a galaxy such as dark matter,
stars, molecular clouds and diffuse interstellar matter (ISM).
Dark matter and stars are treated as collisionless $N$-body
systems. The ISM is numerically described by a smoothed particle
hydrodynamics (SPH) approach for the diffuse (hot) gas and a
sticky particle scheme for the (cool) molecular clouds.
Physical processs such as star formation, stellar death or
condensation and evaporation processes of clouds interacting with
the ISM are described locally. An example
application of the model to a star forming dwarf galaxy will
be shown for comparison with other codes.
Secondly we will discuss new kinds of exotic chemodynamical processes,
as they occur in dense gas-star systems in galactic nuclei, such
as non-standard ``drag''-force interactions, destructive and gas
producing stellar collisions. Their implementation in 1D dynamical
models of galactic nuclei
is presented. Future prospects to generalize these to 3D are
work in progress and will be discussed.

%%%%%%%%%%% Emulate PASA style %%%%%%%%%%%%%%%%%
%\end{abstract}
%%%%%%%%%%% End Emulate PASA style %%%%%%%%%%%%%%%%%
\end{small}
\end{quotation}
]
%%%%%%%%%%% End Emulate PASA style %%%%%%%%%%%%%%%%%

% Place keywords here. Please write all keywords in lower case. PASA uses the
% standard list of subject 
% headings adopted by The Astrophysical Journal and available from URL:
%   http://www.journals.uchicago.edu/ApJ/keywords_text.html

% A formatting command to add space between the author list and the body
% of the paper when printed. This spacing may be changed as desired.
\bigskip

\section{Introduction}

This paper is a short progress report and outlook to show some examples, how
physical and numerical modelling techniques of physical processes in star-gas
systems are improved. The astrophysical aim is to understand nature and formation
of objects, where time scales of the dynamics of the interstellar medium and
dynamical processes in non-dissipative matter (stars or dark matter)
are comparable. The range of objects interesting for us spans
from dwarf galaxies to dense galactic nuclei. This paper is divided into
two main sections, according to applications for dwarf galaxies and galactic
nuclei. For each section a subset of authors is responsible as indicated.

\section{A new chemo-dynamical code and application to dwarf galaxies}
\leftline{\em Authors: Berczik, Hensler, Theis, Spurzem}

\subsection{Introduction}

We present a new 3D chemodynamical code, based on our and other previous
models using smoothed particle hydrodynamics (SPH), including a two phase
interstellar medium consisting of cool clouds and a hot intercloud medium
as an additional feature.
SPH has been invented as a consistent tool to model gasdynamical
systems with gravity by using particles which are subject to non-\-gravi\-tational
forces \citep{Lucy1977, GingoldM1977, M1992} in addition to gravity,
tailored to model in a statistically correct way e.g. pressure or radiation
forces, viscous effect, heating and cooling.
SPH calculations have been applied successfully to
study the formation and evolution of galaxies. Its Lagrangian
nature as well as its easy implementation together with standard
$N$-body codes allows for a simultaneous description of complex
systems consisting of dark matter, gas, and stars
\citep{NW1993, MH1996, CLC1998, TTPCT2000, SYW2001}.
The main features of this SPH variant are: single gas phase,
star formation from SPH particles dependent on the mean mass
density within each individual particle through their free-fall time,
and stellar energy release and mass return to the same particle.
This single-gas phase SPH treatment was successfully applied
to the overall evolution of a Milky Way Galaxy model
\citep{SteinmetzN99, Abadietal03, Ber1999, Ber2000, NakaN03} in the sense that
they could reproduce main structural and chemical signatures
of the global galaxy and of the disk like e.g.
its density profile and metallicity gradient (see also Nakasato, this volume).

In our (multi-phase gas) code we use a two component gas
description of the ISM \citep{TBH1992, SHT1997}. The basic idea
is to add a cold (10$^2$ - 10$^4$ K) cloudy component to the
smooth and hot gas (10$^4$ - 10$^7$ K) described by SPH. The cold
clumps are modeled as $N$-body particles with some ``viscosity''
\citep{TH1993}. This ``viscosity'' models the processes of the
cloud-cloud collisions and also a drag force between clouds and hot
gas component is implemented. The cloudy component interacts with the
surrounding hot gas also via condensation and evaporation
processes \citep{CMcKO1981, KTH1998}. See for a similar
approach also \cite{Harfstetal03} and Harfst (this volume)
and for another 3D chemodynamical model on galaxy formation which is not
based on SPH, but on a mesh-based approach \cite{SamlandG03}, Samland (this volume).
Note also another recent multi-phase model by \cite{SemelinC02},
which globally resembles our approach, but in detail different approximations
of physical processes are used. We would provide a more detailed comparisons
of their and our models elsewhere \citep{Ber2004}.

In the following list we summarize the ingredients of our present model:

\begin{itemize}
\item Hot Gas, treated by SPH, metallicity dependent cooling.
\item Cloud System, sticky particles, using Larson's $M$-$R$ relation.
\item Evaporation of cloud material via thermal conduction in hot medium,
condensation onto clouds via cooling of hot medium. Exchange of energy
and momentum due to this between the components.
\item Ram pressure momentum exchange between clouds moving with relative
velocity to hot medium.
\item Simple star formation prescription out of cold clouds (Schmidt's law),
with delay of star formation due to cloud's collapse time to allow for self-regulation.
%
% Sorry Peter reminded me that this is not yet implemented. RS
%
%\item Cloud formation via thermal cooling of hot medium if critical threshold
%density is exceeded.
%
\item Stellar particles each represent a single stellar population (SSP),
metallicity dependent Padova stellar lifetimes used.
\item Supernovae of type I and II, stellar winds and planetary nebulae feed mass and energy 
back into both the cloud system and the hot interstellar medium.
\end{itemize}

We use a 2nd order two step Runge-Kutta-Fehlberg predictor-corrector scheme, moving the
particles according to the non-\-gravi\-tational forces (SPH, interactions as listed
above) and gravitational forces of all other particles.
To calculate the self gravity we use the {\tt GRAPE5}
computer system at the Astronomical Data Analysis Center of the
National Astronomical Observatory, Japan. A
more detailed description of the board and further links and publications
about {\tt GRAPE} can be found in \\
{\tt http://grape.astron.s.u-tokyo.ac.jp/grape/}

%--------------------------------------%
\begin{figure}
\centering
 \psfig{file=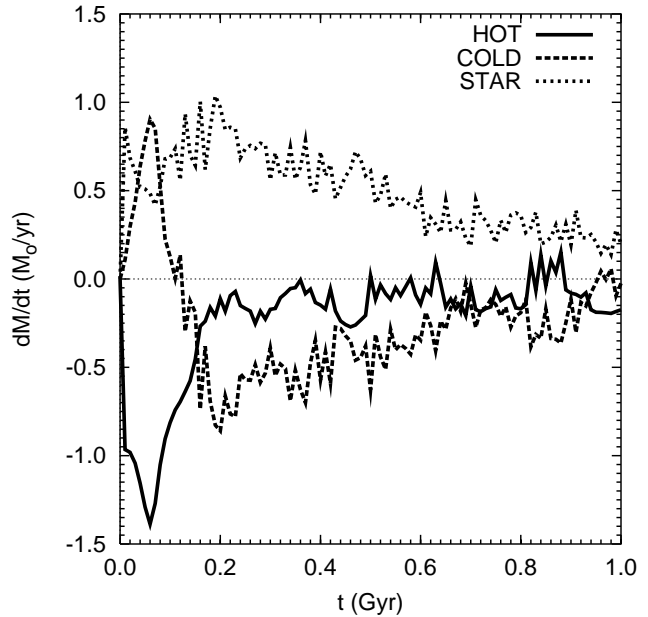,width=\hsize}
  \caption{The temporal evolution of the mass exchange rate for
  the different components of the model galaxy.}
  \label{dmdt-t}
\end{figure}
%--------------------------------------%

\subsection{First Models of Dwarf Galaxies}

We choose the dwarf galaxy as an interesting astrophysical
object to apply our new code,
because in this case even with a relatively ``small'' number of
cold ``clouds'' ($\sim$ 10$^4$) we achieve the required physical
resolution for a realistic description of individual molecular
clouds ($\sim$ 10$^5$ M$_\odot$) as a separate ``cold'' particle.
In the simulation we use N$_{\rm hot}$ = 10$^4$ SPH and N$_{\rm
cold}$ = 10$^4$ ``cold'' particles. After 1~Gyr more then $10^4$
additional ``stellar'' particles are created.

We follow the evolution of an
isolated star forming dwarf galaxy. The initial total gas content
of our dwarf galaxy is $2 \cdot 10^9$ M$_\odot$ (80 \% ``cold''
+ 20 \% ``hot'') which is placed inside a fixed dark matter halo
with parameters $~r_0~=~2$~kpc and $~\rho_0~=~0.075$
M$_\odot$/pc$^3$ \citep{Bur1995}.

With these parameters the dark matter mass inside the initial
distribution of gas (20~kpc) is $\simeq~2 \cdot 10^{10}$
M$_\odot$. The initial temperatures for the cold gas were set
to 10$^3$ K, for the hot gas to 10$^5$ K. For the initial gas
distribution (``cold'' and ``hot'') we use a Plummer-Kuzmin disk with parameters
$~a~=~0.1$~kpc and $~b~=~2$~kpc \citep{MN1975}. The gas initially
rotates in centrifugal equilibrium (in the total ``dm'' + ``gas''
gravitational field) around the z-axis.

Our model first exhibits a strong collapse initiated by
cooling, cloud formation, and subsequently star formation
sets in. In Fig.~\ref{dmdt-t} the growth or loss rates of the different
components are shown. The star formation rate (SFR) peaks to a value of
1 M$_\odot$yr$^{-1}$ after 200 Myrs. Afterwards it drops down to
0.2 M$_\odot$yr$^{-1}$ within several hundred Myrs.
After 1 Gyr the stellar mass has already reached $5 \cdot 10^8$
M$_\odot$. 

%--------------------------------------%
\begin{figure}
\centering
 \psfig{file=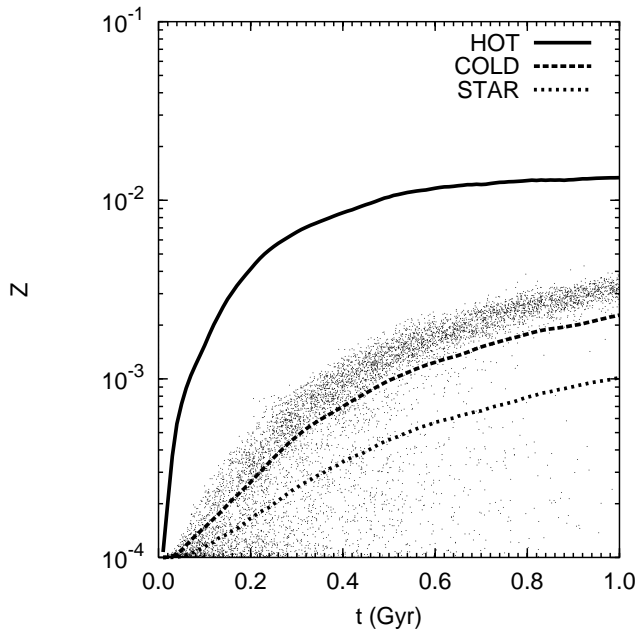,width=\hsize}
  \caption{Temporal evolution of the total metallicity for the
 different components. Individual
  metallicities of newly born stars are marked by dots.}
  \label{z-t}
\end{figure}
%--------------------------------------%

%--------------------------------------%
\begin{figure}
\centering
 \psfig{file=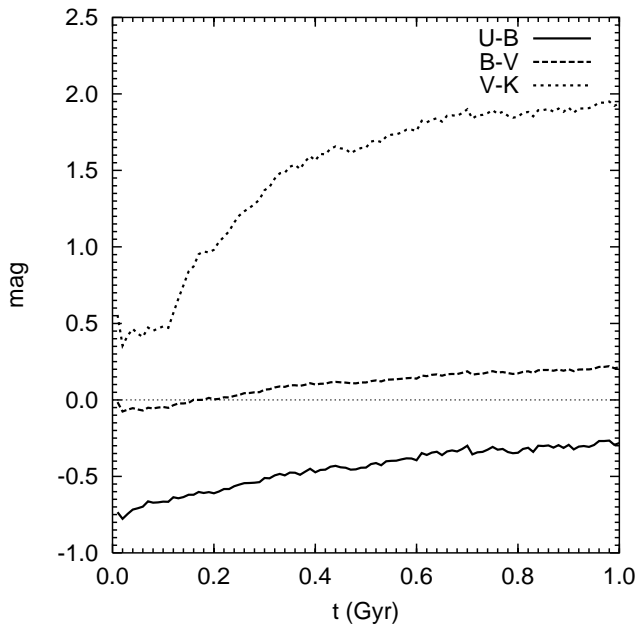,width=\hsize}
  \caption{Temporal evolution of the model galaxy color indices.}
  \label{ci-t}
\end{figure}
%--------------------------------------%

The metal content of the diffuse gas and the clouds differs
significantly over the whole integration time (see Fig.~\ref{z-t}).
Due to SNII and SNIa events the metallicity
of the hot phase exceeds that of the clouds by almost one order of
magnitude. The clouds mainly get their metals by condensation of
the hot phase. This shows that the two phases of the interstellar
medium exhibit dynamically and chemically a different behaviour and
thus a correct physical treatment like ours is required for reliable
modelling.

In Fig.~\ref{ci-t} we present the
evolution of the color indices in our model galaxy as an example
what kind of data can be constructed for comparison with observational
data. Our SSP model provides spectral information via six photometric
channels from each star particle (representing its own SSP); the reader
interested in more details about this or other features of our model
is referred to  
\citep{Ber1999, Ber2000, Ber2003, Ber2004}.

\section{Dense Star-Gas Systems in Galactic Nuclei}
\leftline{\em Authors: Amaro-Seoane, Freitag, Just, Spurzem}

At the time of the first collapse and star formation epoch in galaxy formation
large amounts of gas will reach the galactic centre \citep{EisensteinL95,ZoltanL01}.
They may create a huge outburst of star formation, and in part be responsible for
the ultraluminous IR galaxies in the young universe.
In addition to that, once stars have been formed,
very large amounts of additional gas will be liberated by disruptive stellar
collisions \citep{FreitagB02}. In some cases of massive galactic nuclei
gas production rates as large as 50 M$_\odot$/yr have been observed in the
numerical models. 
Present stellar dynamical models based on 
Monte Carlo or direct solutions of the Fokker-Planck equation do not include star-gas
interactions. Thus they cannot follow the co-evolution of such a system,   
as is done in classical chemodynamical galaxy simulations.
There is one exception, the so-called gaseous
model of stellar dynamics \citep{LouisSp91,Langbeinetal90,Amaroetal04}, which at least
in principle could be coupled in an easy way to simulate joint gas and stellar
dynamics. The reader interested in more detail and more references on the
exotic star-gas interaction processes is referred to \cite{Langbeinetal90}, where
the terms are given in detail, and the history of previous literature on the
subject is presented.

%--------------------------------------%
\begin{figure}
 \psfig{file=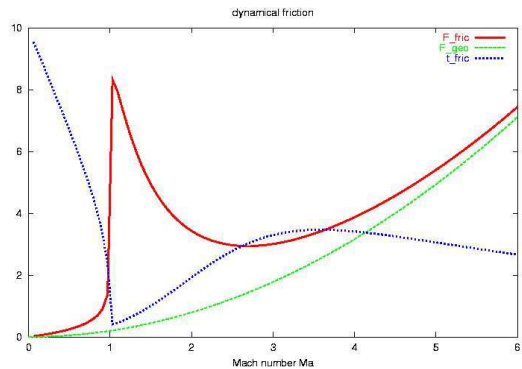,width=1.00\hsize}
  \caption{``Drag'' force on a star moving through interstellar medium $F_{\rm geo}$ with
Mach number Ma from a standard estimate using the geometrical stellar cross section and
a ram pressure approach (see main text), compared with a more realistic dissipative interaction
force $F_{\rm fric}$ obtained from a detailed analysis of fluctuations in the ISM induced by a star
moving through (see main text for citations and more explanations). The dotted curve is just
inversely proportional to $F_{\rm fric}$, since it is plotting the time scale connected to it.}
  \label{drag}
\end{figure}
%--------------------------------------%

Recently it has been pointed out that the formation
of clusters of intermediate mass black holes \citep{Ebisuzakietal01} would be
a possible way to create massive black holes in centres of star clusters or
galaxies. This is nicely suggested also by new Chandra images of M82 by Fabbiano et al.,
see for a reference \\
{\tt http://chandra.harvard.edu/photo/cycle1/0094true/}\\
and recent stellar dynamical Monte Carlo models including stellar evolution and mergers 
have strengthened this idea \citep{Rasioetal:03,Guerkanetal:03}.
In order to determine what initial and boundary conditions (e.g. from large
scale galaxy formation) lead to the formation of supermassive black holes and
what determines their further growth by star and gas accretion, and what will be
the spectrophotometric appearance of such dense nuclei, we need to extend the
classical chemodynamics into the regime of exotic processes in galactic nuclei.
These are gas production by stellar collisions, star-gas drag under special
conditions (covering very large ranges of Mach numbers), energy transport in the
presence of energetic radiation which contributes to the hydrodynamic pressure
significantly. 
The assumption of spherical symmetry on which the Monte Carlo and gas methods
rely becomes highly questionable in the close vicinity of the black hole.
One very recent remarkable attempt 
of \cite{BrommL03} uses an SPH ansatz, but is not able to follow the multi-phase
dynamics of the interstellar matter, as we can do with our CD-SPH model presented in
the first section (see also S. Harfst, this volume, for a similar ansatz). One example
of the new and complex phenomena is presented in Fig.~\ref{drag}, which shows the
star-gas interaction. Fig.~\ref{drag} shows a standard ram pressure force for a star moving
supersonically in an ambient medium, as e.g. discussed in \cite{Bisnovatyi-KoganS72},
where the ``drag'' force is proportional to the square of velocity (Mach number) and
linearly dependent on the geometrical stellar cross section (strictly, since the velocity
dependence is quadratic, this is not a drag force, but it is often called so, therefore
we put ``drag'' in quotation marks to remind the reader about this ambiguity).
The standard ``drag'' force is compared with a non-standard one, which is valid also
in the subsonic and transition regime (Ma $\approx 1$). The underlying formalism used
is the analysis of fluctuations induced by a star moving through an ambient ISM and
the feedback force they provide on the star's motion. For small Mach numbers Ma$>1$ the
resulting force is large and related to a kind of dynamical friction process, while
for highly supersonic motion we reach the asymptotic limit of the standard formula. 
In the subsonic case the interaction force drops sharply, but still differs from the
standard formula. The basic method of the analysis of such kind of dynamical ``drag''
between stars and the ISM is described in \cite{Justetal86}. The application to galactic
nuclei will be presented in ongoing work (Just, 2004, in prep.).
In galactic centres we expect that all ranges from subsonic to hypersonic
(Mach numbers of a few thousand) will be realised.

\section*{Acknowledgments}

This work has been supported by Sonderforschungsbereich (SFB) 439 ``Galaxies
in the Young Universe´´ at the Univ. of Heidelberg in sub-projects A5 (RS, AJ) and
B5 (RS, PB). PB acknowledges kind hospitality at the Univ. of Kiel and to
G. Hensler, Ch. Theis and collaborators, where part of this work has been
performed. Numerical models have been computed with the {\tt GRAPE5}            
 system at the Astronomical Data Analysis Center of the
 National Astronomical Observatory, Japan.

%\section*{References}

% for a journal article, or

%\reference Author, A.B. \and Anotherauthor, C. D. 1990, in This is a Book %Title, ed. C. D. Editor, (City: Publisher Name), 437

% for a book.

%\reference REFERENCE1
%\reference REFERENCE2
% Add as many references as required.

\end{document}